\pdfoutput=1
\documentclass[12pt]{article}

\font\grande=cmr9.5 scaled \magstep4
\font\medio=cmr9.5 scaled \magstep2
\outer\def\beginsection#1\par{\medbreak\bigskip
      \message{#1}\leftline{\bf#1}\nobreak\medskip
\vskip-\parskip
      \noindent}

\begin{document}
\bibliographystyle {unsrt}

\titlepage

\begin{flushright}
CERN-TH/2016-194
\end{flushright}

\vspace{10mm}
\begin{center}
{\grande Stringy bounces and gradient instabilities}\\
\vspace{1.5cm}
 Massimo Giovannini
 \footnote{Electronic address: massimo.giovannini@cern.ch}\\
\vspace{1cm}
{{\sl Theory Division, CERN, 1211 Geneva 23, Switzerland }}\\
\vspace{0.5cm}
{{\sl INFN, Section of Milan-Bicocca, 20126 Milan, Italy}}
\vspace*{0.5cm}
\end{center}

\vskip 0.5cm
\centerline{\medio  Abstract}
Bouncing solutions are obtained from a generally covariant action characterized by a potential 
which is a nonlocal functional of the dilaton field at two separated space-time points. 
Gradient instabilities are shown to arise in this context but they are argued to be nongeneric. 
After performing a gauge-invariant and frame-invariant derivation of the evolution equations 
of the fluctuations, a heuristic criterion for the avoidance of pathological instabilities is 
proposed and corroborated by a number of explicit examples that turn out to be 
compatible with a quasi-flat spectrum of curvature inhomogeneities for large wavelengths.
\vskip 0.5cm

\noindent

\vspace{5mm}
\vfill
\newpage
\renewcommand{\theequation}{1.\arabic{equation}}
\setcounter{equation}{0}
\section{Introduction}
\label{sec1}
The temperature and polarization anisotropies of the Cosmic Microwave Background can be 
successfully reproduced by assuming that the initial conditions of the Einstein-Boltzmann hierarchy are 
predominantly adiabatic and Gaussian \cite{WMAP1,WMAP2}. Inflationary models are consistent with 
the presence of a dominant adiabatic mode but the smallness of the tensor to scalar ratio calls 
for a plateau-like potential in the Einstein frame implying a minute 
energy density of the inflaton in Planck units (typically  $\rho_{inf} = {\mathcal O}(10^{-12}) \, M_{P}^4$).
A very plausible chain of arguments \cite{inf1} stipulates that 
the kinetic energy of the inflaton and the spatial curvature must 
 then be comparable with the inflaton potential if we do not want 
the kinetic energy and the spatial curvature to dominate even before inflation starts.
This occurrence can be quantitatively scrutinized by following the evolution of the spatial gradients 
\cite{inf2} during the preinflationary phase. A complete theory of the initial conditions 
should account for the emergence of the observable Universe from a sufficiently 
generic set of initial data. It might well be, however, that inflation should be primarily regarded 
as a model of the spectral indices rather than a theory of the initial data. 

Since bouncing models more often than not lead to a small tensor components over large 
distance scales, they have been intensively investigated in the last twenty 
years with various independent motivations ranging from string inspired models to different classes 
of scenarios concocted in the framework of effective theories 
(see, for instance, \cite{boun1} for some reviews on the subject).
The current versions of bouncing models have many virtues but they 
also have various well known problems which have been scrutinized in the past 
and are currently under active consideration \cite{GI1,GI2}. 
Some of the current approaches even renounce general covariance and 
often impose ad hoc symmetries (like the shift symmetry). 

An interesting class of bouncing models can be obtained, in the low-curvature regime, by the contribution 
of a nonlocal, though generally covariant, dilaton potential \cite{nonloc1}. Even if the potential 
is nonlocal, the corresponding evolution equations of the background and of the 
corresponding fluctuations are perfectly local in time. Bouncing solutions have been 
studied in this framework. The production of massless quanta (e.g. gauge bosons) could heat
 the background, stabilize the dilaton and eventually provide an exit to radiation 
(see, in particular, first paper of \cite{nonloc1}). The idea that radiation 
can arise from the backreaction of the produced quanta goes back to the 
seminal contributions of various authors \cite{BR} (see, in particular, the analyses
of Parker and Ford). Bouncing solutions may also arise in the context of double field theory \cite{double1} which 
was firstly proposed to realize $T$-duality explicitly at the level of component fields of closed string field theory; 
earlier contributions along this direction can be found in \cite{double2}. Since the nonlocal potential of \cite{nonloc1} depends on a $T$-duality invariant combination, the corresponding solutions can also be interpreted in a 
double field theory context \cite{double3}.

Various bouncing models may experience the so called gradient instability stipulating that 
either the scalar or the tensor modes of the geometry inherit an imaginary sound speed. 
The corresponding fluctuations are then exponentially unstable for sufficiently small wavelengths.  To  cure this pathology non-covariant terms
are often added to the action of the fluctuations with the purpose of modifying 
 the effective sound speed without disturbing too much the evolution of the background. 
In the present investigation, after discussing in detail the derivation of the scalar sound speed, it will be 
demonstrated that gradient instabilities take place also in the context of bouncing models induced by a nonlocal dilaton potential. 
Unlike other modes, however, these instabilities do not arise ubiquitously for any form 
of the dilaton potential and are, in this sense, nongeneric. We provide examples of 
semi-realistic backgrounds where the instabilities are tamed and the spectrum of the scalar modes turns 
out to be quasi-flat. 

We remark that while the scalar modes of the geometry 
may evolve differently in the Einstein and in the string frames, 
the curvature perturbations on comoving orthogonal 
hypersurfaces are both gauge-invariant and frame-invariant. The corresponding
 evolution equations inherit an effective sound speed.
 Such a terminology is justified since the pivotal equation derived in this 
context coincides, formally, with the equation derived long ago by 
Lukash \cite{lukash} and describing the excitations of a relativistic and irrotational 
fluid in a Friedmann-Robertson-Walker background. Note, however that in the case of Ref. \cite{lukash} the 
sound speed is the square root of the ratio between the derivatives of the pressure and on the 
energy density of the relativistic fluid, as it should be by definition of the sound speed; 
in the present case, however, the effective sound speed will ultimately be a functional of the potential. 
A valid help for this analysis is represented by the uniform dilaton gauge
which has been discussed in various related contexts \cite{hw1,hw2}. 
In this gauge the dilaton remains unperturbed, the curvature perturbations 
coincide with the longitudinal degrees of freedom of the perturbed metric and 
the scalar modes are automatically frame-invariant.
 
The present paper is organized as follows. In section \ref{sec2}
 the bouncing models induced by nonlocal dilaton potentials are introduced. In section \ref{sec3} we derive explicitly 
the scalar sound speed and discuss the problem of the gradient instability.
The genericness of the gradient instability is scrutinized in section \ref{sec4} with the 
aim of constructing  solutions where this pathology does not arise. A number of potential 
drawbacks are anyway suggested in the last part of section \ref{sec4}. 
Finally section \ref{sec5} contains the concluding remarks.  
Various results have been presented and summarized in a self-contained 
perspective in the appendix \ref{APPA}.

\renewcommand{\theequation}{2.\arabic{equation}}
\setcounter{equation}{0}
\section{Nonlocal potentials and bouncing solutions}
\label{sec2}
A smooth bouncing transition at low curvatures can be achieved in the framework of the 
following generally covariant action in four space-time dimensions \cite{nonloc1}:
\begin{equation}
S= - \frac{1}{\lambda_{s}^{2} }\int d^{4} x \sqrt{- g} e^{- \varphi} \biggl[ R + 
g^{\alpha\beta} \nabla_{\alpha} \varphi \nabla_{\beta} \varphi + V\biggr],
\label{action1}
\end{equation}
where we have that $V= V(e^{-\overline{\varphi}})$ and 
\begin{equation}
e^{-\overline{\varphi}(x)} = \frac{1}{\lambda_{s}^{3}} \int d^{4} w \sqrt{-g(w)} e^{- \varphi(w)} 
\sqrt{g^{\alpha\beta} \partial_{\alpha} \varphi (w)\partial_{\beta} \varphi(w)} \delta[ \varphi(x) - \varphi(w)\biggr].
\label{action2}
\end{equation}
For immediate convenience we also define the following pair of integrals:
\begin{eqnarray}
&&{\mathcal I}_1 = \frac{1}{\lambda_{s}^{3}} \int d^{4} w \sqrt{-g(w)} V' (e^{- \overline{\varphi}(w)}) \delta[\varphi(x) - \varphi(w)],
\nonumber\\
&&{\mathcal I}_{2} = \frac{1}{\lambda_{s}^{3}} 
\int d^{4} w \sqrt{-g(w)} \sqrt{g^{\alpha\beta} \partial_{\alpha}\varphi(w) \partial_{\beta} \varphi(w)}\delta^{\prime}[\varphi(x) - \varphi(w)],
\label{def}
\end{eqnarray}
with a prime denoting differentiation with respect to the argument\footnote{In Eq. (\ref{eq2}) as well as in  Eqs. (\ref{def}) and (\ref{comb}) the prime denotes a derivation with respect to the argument of the given functional and not the derivative with respect to the conformal time coordinate. The two notations cannot be confused since the conformal time derivative only appears in connection with the explicit form of the equations discussed in Eqs. (\ref{Eone})--(\ref{Ethree}) and in the forthcoming sections.}.  The present analysis can be 
generalized to include antisymmetric tensor fields, gauge fields as well as internal (contracting) extra-dimensions \cite{nonloc2}.These potential contributions will be ignored and we shall focus on the minimal 
four-dimensional  scenario compatible with the presence of adiabatic curvature perturbations.
The variation of the action (\ref{action1}) with respect to $g_{\mu\nu}$ and $ \varphi$ leads, respectively,  to the following pair of equations:
\begin{eqnarray}
&&{\mathcal G}_{\mu\nu}  + \nabla_{\mu} \nabla_{\nu} \varphi + 
\frac{1}{2} g_{\mu\nu} \biggl[  (\partial \varphi)^2 - 2 
g^{\alpha\beta} \nabla_{\alpha} \nabla_{\beta} \varphi - V\biggr] - \frac{e^{-\varphi}}{2} 
\sqrt{ (\partial \varphi)^2} \gamma_{\mu\nu} {\cal I}_{1} = 0,
\label{eq1}\\
&& R + 2 g^{\alpha\beta} \nabla_{\alpha} \nabla_{\beta} \varphi - 
 (\partial \varphi)^2+ V - \frac{\partial V}{\partial\overline{\varphi}} 
+ e^{-\varphi} \frac{\overline{\nabla}^2 \varphi}{\sqrt{ (\partial \varphi)^2}} {\cal I}_1 - e^{-\varphi} V' {\cal I}_{2} =0,
\label{eq2}
\end{eqnarray}
where the shorthand notation $g^{\alpha\beta} \partial_{\alpha}\varphi \partial_{\beta} \varphi= (\partial \varphi)^2 $ has been consistently employed; in Eq. (\ref{eq2}) 
${\mathcal G}_{\mu\nu}$ denotes the Einstein tensor and  $\overline{\nabla}^2$ is defined as:
\begin{equation}
\overline{\nabla}^2 = \gamma^{\mu\nu} \nabla_{\mu}\nabla_{\nu}, \qquad \gamma_{\mu\nu} = g_{\mu\nu} - \frac{\partial_{\mu} \varphi \partial_{\nu} \varphi}{(\partial \varphi)^2}.
\end{equation}
Note that, by definition, we also have the following chain of equalities:
\begin{equation}
\frac{\overline{\nabla}^2 \varphi}{\sqrt{(\partial \varphi)^2}}=
\nabla_\mu \left (g^{\mu\nu}\nabla_\nu \varphi\over \sqrt{(\partial \varphi)^2}\right)=
{1\over \sqrt{-g}}\partial_\mu \left (\sqrt{-g}g^{\mu\nu} \partial_\nu \varphi\over \sqrt{(\partial
\varphi)^2}\right).
\end{equation}
A detailed derivation of Eqs. (\ref{eq1}) and (\ref{eq2}) has been discussed elsewhere \cite{nonloc1} and here we shall only focus on those 
aspects that are germane to the main theme of the analysis. 
By combining Eqs. (\ref{eq1}) and (\ref{eq2})  the Ricci scalar can be eliminated and  the following equation is readily obtained:
\begin{equation}
R_{\mu\nu} + \nabla_{\mu}\nabla_{\nu} \varphi - \frac{1}{2} g_{\mu\nu} \biggl[ \frac{\partial V}{\partial 
\overline{\varphi}} + e^{-\varphi} V' {\mathcal I}_{2} \biggr] + \frac{1}{2} e^{-\varphi} \biggl[  g_{\mu\nu} 
\frac{\overline{\nabla}^2 \varphi}{\sqrt{ (\partial \varphi)^2}} - 
\gamma_{\mu\nu} \sqrt{ (\partial \varphi)^2}\biggr] {\mathcal I}_{1} = 0.
\label{comb}
\end{equation}
In the case of a homogeneous dilaton and for a conformally flat metric of Friedmann-Robertson-Walker type Eqs. (\ref{eq1}) and (\ref{comb}) lead, 
in units $2 \lambda_{s}^2 =1$, to the following system of equations\footnote{The discussion of the gradient instability and of its 
implications can be easily extended to the case $D= d + n +1$ where $d$ and $n$ denote the number of external and internal (i.e. compactified) 
dimensions \cite{nonloc2}. Since this is analysis is not central to the theme of this paper, it will be omitted.}: 
\begin{eqnarray}
&& \dot{\overline{\varphi}}^2 - 3 H_{s}^2 - V = 0,\qquad \dot{H}_{s} = \dot{\overline{\varphi}} H_{s},
\label{b1}\\ 
&& 2 \ddot{\overline{\varphi}} - \dot{\overline{\varphi}}^2 - 3 H_{s}^2 + V - \frac{\partial V}{\partial \overline{\varphi}} =0,
\label{b2}
\end{eqnarray}
where the overdot denotes a derivation with respect to the cosmic time coordinate $t$; as usual $H= \dot{a}_{s}/a_{s}$ and, in the homogeneous 
limit $\overline{\varphi} = \varphi - 3 \ln{a_{s}}$.  Equations (\ref{b1})--(\ref{b2}) hold in the string frame\footnote{The evolution of the background (and of its fluctuations) can be described either in the string  
frame (where the dilaton and the Ricci scalar are explicitly coupled) or in the Einstein frame. A self-contained discussion of the relation
 between the two conformally related frames can be found in appendix \ref{APPA}.}.
For the discussion of the fluctuations it is practical to write Eqs. (\ref{b1})--(\ref{b2}) in the conformal time coordinate $\tau$ 
and, eventually, in the Einstein frame. Unlike the cosmic time coordinate the conformal time is frame-invariant (i.e. $\tau_{s} = \tau_{e}=\tau$). The equations for the background in the Einstein frame are\footnote{Equations (\ref{Eone})--(\ref{Ethree}) are written in natural gravitational units $16\pi G =1$; similarly Eqs. (\ref{b1})--(\ref{b2}) 
have been written in natural string units $2 \lambda_{s}^2$=1.}:
\begin{eqnarray}
&& 6  {\mathcal H}_{e}^2 = \frac{1}{2} { \varphi'}^2  + e^{\varphi} a_{e}^2 V,
\label{Eone}\\
&&  4 {\mathcal H}_{e}^{\prime} + 2 {\mathcal H}_{e}^2 = -\biggl(\frac{{\varphi}^{\prime\,2}}{2} - e^{\varphi} \,a_{e}^2 \, V \biggr) - e^{\varphi} a_{e}^2 \frac{\partial V}{\partial \overline{\varphi}},
\label{Etwo}\\
&& \varphi^{\prime\prime} + 2 {\mathcal H}_{e} \varphi^{\prime} + e^{\varphi} a_{e}^2 \biggl( V - \frac{1}{2}  \frac{\partial V}{\partial \overline{\varphi}}\biggr)=0.
\label{Ethree}
\end{eqnarray}
 Equations (\ref{Eone})--(\ref{Ethree}) can be simply derived by first writing Eqs. (\ref{b1}) and (\ref{b2})  in terms of the conformal time $\tau$ and by then transforming the result to the Einstein frame (see, in particular, the last part of the appendix \ref{APPA}). 
In what follows we shall use indifferently either Eqs. (\ref{b1})--(\ref{b2}) or (\ref{Eone})--(\ref{Ethree}). 
Note finally that the energy density and pressure in the Einstein frame description are simply given by:
\begin{equation}
\rho_{e} = \frac{{\varphi '}^2}{ 2 a_{e}^2} + e^{\varphi} V, \qquad p_{e} = \frac{{\varphi '}^2}{2 a_{e}^2} - e^{\varphi} V + 
e^{\varphi} \frac{\partial V}{\partial \overline{\varphi}}.
\label{pr}
\end{equation}
In terms of $\rho_{e}$ and $p_{e}$ Eq. (\ref{Ethree}) simply becomes $\rho_{e}' + 3 {\cal H}_{e} (\rho_{e} + p_{e}) =0$, as expected.

As discussed in the introduction, even if the potential is nonlocal in field space the evolution equations of the background are local in time \cite{nonloc1}.
The effect of the nonlocal modification of the action (\ref{action1}) is however apparent in the modification of the equations 
of motion in a way that makes possible the bouncing solution. To appreciate this important point, let us 
suppose, for a moment, that the potential term in Eq. (\ref{action1}) is just a local function of $\varphi$, i.e. $V= V(\varphi)$ and not 
$V= V(\overline{\varphi})$, as Eq. (\ref{action2}) stipulates. To avoid confusions between the two situations let us write, for notational convenience, 
that the local potential corresponds to $W(\varphi)$. In this case the evolution equations can be immediately obtained 
either in the string frame or in the Einstein frame. To make the comparison more clear let us therefore write the analog of Eqs. (\ref{Eone})--(\ref{Ethree})
when the action is given by Eq. (\ref{action1}) but the potential is $W(\varphi)$
\begin{eqnarray}
&& 6  {\mathcal H}_{e}^2 = \frac{1}{2} { \varphi'}^2  + e^{\varphi} a_{e}^2 W,
\label{Eonea}\\
&&  4 {\mathcal H}_{e}^{\prime} + 2 {\mathcal H}_{e}^2 = -\biggl(\frac{{\varphi}^{\prime\,2}}{2} - e^{\varphi} \,a_{e}^2 \, W \biggr),
\label{Etwoa}\\
&& \varphi^{\prime\prime} + 2 {\mathcal H}_{e} \varphi^{\prime} + e^{\varphi} a_{e}^2 \biggl( W + \frac{\partial W}{\partial \overline{\varphi}}\biggr)=0.
\label{Ethreea}
\end{eqnarray}
Note that in Eq. (\ref{Ethreea}) we also have a term going as $W$; this can be understood by 
conformally transforming the action (\ref{action1}) in the Einstein frame where the potential is $e^{\varphi} W(\varphi)$. 
If we now subtract Eq. (\ref{Etwoa}) from Eq. (\ref{Eonea}) we simply obtain 
\begin{equation}
 {\mathcal H}_{e}^2 - {\mathcal H}_{e}^{\prime} = { \varphi'}^2 ,
 \label{sub1}
 \end{equation}
 Since ${ \varphi'}^2$ is always positive semidefinite there is no way of obtaining bouncing solutions in this context. Let us 
 now do the same exercise with Eqs. (\ref{Eone})--(\ref{Ethree}); more specifically let us subtract 
 Eq. (\ref{Etwo}) from Eq. (\ref{Eone}). The result of this manipulation is 
\begin{equation}
 {\mathcal H}_{e}^2 - {\mathcal H}_{e}^{\prime} = { \varphi'}^2 +   e^{\varphi} a_{e}^2 \frac{\partial V}{\partial \overline{\varphi}},
 \label{sub2}
 \end{equation}
 showing that in this case, depending on the sign of the derivative of $V$ with respect to $\overline{\varphi}$,
 bouncing solutions become possible. All in all we can therefore conclude that even if the evolution 
 equations (\ref{Eone})--(\ref{Ethree}) are local in time, the nonlocality of the potential which depends on the shifted dilaton $\overline{\varphi}$ is reflected 
 in a substantially different form of the equations which cannot be mimicked 
 by a potential term depending only on $\varphi$.
 
 The essential features of Eq. (\ref{sub2}) are not affected by the isotropy 
 of the metric. Indeed when the potential depends on the shifted dilaton the solutions can be 
 anisotropic both in $3+1$ dimensions as well as in higher dimensions \cite{nonloc1}. In particular $10$
 dimensional solutions have been used to discuss the evolution of the vector modes of the geometry (see, in this 
 respect, the first paper of Ref. \cite{nonloc1}). Clearly depending on the frame where the solutions 
 are described the features of the solutions will be slightly different. In the anisotropic case, however, the solutions 
 will be of Kasner type: some of the dimensions will expand and some will contract. 
 
 Let us finally remark, as we close the section, that the potentials depending on the shifted dilaton 
 may be interpreted as string loop corrections preserving the $T$-duality symmetry \cite{nonloc1}.
 This interpretation is particularly intriguing in the light of the potential interpretation 
 of the solutions in the light of double field theory \cite{double1,double2,double3}.  Having said this, 
 not all the potentials depending on $\overline{{\varphi}}$ induce bouncing 
 solutions: this method of regularizing the bouncing solutions remains after all 
 heuristic. This observation is of course a source of concern but it should be contrasted 
 with the effective approaches to bouncing solutions  \cite{GI1,GI2}.

\renewcommand{\theequation}{3.\arabic{equation}}
\setcounter{equation}{0}
\section{Gradient instabilities}
\label{sec3}
In the present framework the gradient instabilities do not affect the evolution of the tensor modes  
while the scalar modes inherit an effective sound speed which may get 
imaginary. Denoting, in general terms, by $c_{t}^2$ and $c_{s}^2$ the sound speeds of the tensor 
and of the scalar modes of the geometry we have that a gradient instability 
is said to arise when either $c_{t}^2<0$ or when $c_{s}^2<0$. 
The tensor modes of the geometry are both gauge-invariant 
(i.e. invariant under infinitesimal coordinate transformations) and frame-invariant 
(i.e. invariant for the transition from the string to the Einstein frame).
The scalar modes are not automatically frame-invariant but, as we shall 
discuss the gauge-invariant curvature perturbations are also 
frame-invariant (see also appendix \ref{APPA}). Finally, as already mentioned, 
the conformal time coordinate (unlike the cosmic time parametrization) is frame-invariant.

The evolution equations of the fluctuations in the string frame can be obtained by perturbing to first-order Eq. (\ref{eq1}). Denoting 
by $\delta g_{\mu\nu}$ the total fluctuation of the metric, the tensor, scalar and vector modes will be:
\begin{equation}
\delta g_{\mu\nu} = \delta_{t} g_{\mu\nu} + \delta_{s} g_{\mu\nu} + \delta_{v} g_{\mu\nu},
\end{equation}
where the subscripts remind, respectively, of the tensor, scalar and vector perturbations. 
We shall discuss hereunder the tensor and the scalar modes of the bounce. 
As a consequence of the bounce the vector zero modes may increase, but their fate depends upon the dynamics. This 
problem has been specifically addressed (see second paper in Ref. \cite{nonloc1}) and the result can be summarized by saying that 
for non-singular bounces in four-dimensions the growing mode customarily present during the contracting phase matches with the decaying mode after the bounce\footnote{In the multidimensional case the situation becomes increasingly interesting \cite{nonloc2}: the vector modes are more numerous than in the four-dimensional case and their quantum mechanical fluctuations can be amplified (see first paper of \cite{nonloc1}). We shall therefore neglect the dynamics of the vector modes since they are not central to the theme of discussion which is bound to four space-time dimensions.}. 

\subsection{Evolution of the tensor modes}
In spite of the specific background solution deduced from the action (\ref{action1}) the tensor modes do not experience any sort of gradient instability. To demonstrate this statement the tensor fluctuations can be defined, in 
the string frame, as: 
\begin{equation}
\delta_{t} g_{i j} = - a_{s}^2 h_{ij},\qquad  \partial_{i} h^{ij} = h_{i}^{i} =0,
\label{T0}
\end{equation}
where $a_{s}$ denotes the scale factor in the string frame metric and $\delta_{t}$ stands for  the 
tensor fluctuation of the corresponding quantity. Let us therefore perturb Eq. (\ref{eq1}) when the indices are mixed (i.e. one covariant and the other contravariant):
\begin{eqnarray}
&& \delta_{t} {\mathcal G}_{\mu}^{\nu} + (\nabla_{\alpha} \nabla_{\mu}\varphi) \,  \delta_{t} g^{\nu\alpha} 
- (g^{\nu\alpha} \partial_{\sigma} \varphi) \,  \delta_{t} \Gamma_{\alpha\mu}^{\sigma} =0.
\label{PERT1}
\end{eqnarray}
All the quantities not preceded by $\delta_{t}$ in Eq. (\ref{PERT2}) must be understood as evaluated on the background.
Recalling Eq. (\ref{T0}) the explicit fluctuations of the Einstein tensor and of the Christoffel connections can be 
readily computed; the equation for $h_{ij}$ becomes then:
\begin{equation}
 h_{ij}^{\prime\prime} - ({\overline{\varphi}}^{\prime} + {\mathcal H}_{s}) h_{ij}^{\prime} - \nabla^2 h_{ij}=0,\qquad {\mathcal H}_{s} = \frac{a_{s}^{\prime}}{a_{s}},
\label{T1}
\end{equation}
where the prime denotes the derivation with respect to the conformal time coordinate; the relation between ${\mathcal H}_{s}$ and $H_{s}$ (defined 
after Eqs. (\ref{b1})--(\ref{b2})) is given as usual by  ${\mathcal H}_{s} =a_{s} H_{s}$. The equations for the tensors in the Einstein frame are given by:
\begin{equation}
 h_{ij}^{\prime\prime} + 2 {\mathcal H}_{e} h_{ij}^{\prime} - \nabla^2 h_{ij}=0,\qquad {\mathcal H}_{e} = \frac{a_{e}^{\prime}}{a_{e}}.
\label{T2}
\end{equation}
Equation (\ref{T2}) can be obtained from Eq. (\ref{T1}) by appreciating that the tensor amplitudes are frame-invariant 
(see appendix \ref{APPA} for further details). All in all Eqs. (\ref{T1}) and (\ref{T2}) show that there are no problems with the gradient instability in the case of the tensor modes which are automatically gauge-invariant and frame-invariant. This conclusion should be contrasted with the remark mentioned at 
the beginning of this section where, in general terms, it has been said that the tensor modes may inherit an effective sound speed. 
This can happen, for instance, if the bounces are described using effective field theory methods \cite{GI1,GI2} analog to the one employed in the context 
of inflationary modes \cite{wein}. 

\subsection{Evolution of the scalar modes}
The scalar fluctuations of the geometry are given by:
\begin{equation} 
\delta_{s} g_{00} = 2 a_{s}^2 \phi,\qquad \delta_{s} g^{(s)}_{i j} 
= 2 a_{s}^2 ( \psi \delta_{ij} - \partial_{i}\partial_{j } E),\qquad \delta_{s} g_{0i} = - a_{s}^2 \partial_{i} B,
\label{S1}
\end{equation}
where $\delta_{s}$ denotes  the scalar fluctuation of the corresponding quantity.
We fix the coordinate system by setting to zero the perturbation of the dilaton field
and of the off-diagonal fluctuations of the metric.  This 
is often referred to as the uniform field gauge \cite{hw1,hw2} and it is particularly practical in the present context. 
Since the dilaton and its fluctuation are frame-invariant we can denote by $\delta_{s}\varphi = \chi$ 
the common value of 
the dilaton fluctuation either in the Einstein or in the string frame. The uniform field  gauge stipulates that $\chi=0$ and $B=0$. Clearly
 for an infinitesimal coordinate transformation\footnote{The following notations shall be employed: $\epsilon^{\mu} = (\epsilon^{0}, \, \epsilon^{i})$ implying $\epsilon_{\mu} = a_{s}^2(\epsilon_{0}, - \epsilon_{i})$ with $\epsilon_{i} = \partial_{i} \epsilon$. }
parametrized as $x^{\mu} \to \widetilde{x}^{\mu} = x^{\mu} + \epsilon^{\mu}$ the fluctuation of a rank-two tensor in four-dimensions (like $\delta_{s} g_{\mu\nu}$) change according to the Lie derivative in the direction $\epsilon^{\mu}$. In explicit terms we will have that 
\begin{equation}
B \to \widetilde{B} =B + \epsilon^{0} - \epsilon^{\prime}, \qquad \chi \to \widetilde{\chi} =  \chi - \varphi' \epsilon^{0},
\label{S1a}
\end{equation}
where we also reported, for immediate convenience, the gauge transformation of $\chi$.
If we start from a generic gauge we can arrive at the uniform dilaton gauge by setting  $\epsilon^{0} = \chi/\varphi'$. 
Furthermore by setting $\tilde{B}=0$ in Eq. (\ref{S1a}) the value of $\epsilon$ can be determined and it is: 
\begin{equation}
\epsilon= \int ( B + \chi/\varphi') d\tau + c_1,
\label{int}
\end{equation}
where $c_{1}$ is an integration constant which does not depend on the conformal time coordinate. Since gauge freedom is not completely fixed the evolution equations 
of the scalar modes (see below) depend on $E'$ (and not on $E$) which is related to the gauge-invariant 
dilaton fluctuation in the uniform field gauge\footnote{Indeed the gauge-invariant fluctuation of the dilaton (see also appendix \ref{APPA}) 
is frame-invariant and it is give by $X= \chi + \varphi^{\prime} (B-E^{\prime})$; in the uniform dilaton gauge $X = - \varphi^{\prime} E^{\prime}$.}. In the case of the scalar modes of the geometry the perturbed version of Eq. (\ref{eq1}) can be written in a more explicit form as: 
\begin{eqnarray}
&& \delta_{s} {\mathcal G}_{\mu}^{\nu} + \delta_{s} g^{\nu\alpha} \biggl[ \partial_{\alpha}\partial_{\mu} \varphi - \Gamma_{\alpha\mu}^{\sigma} \partial_{\sigma} \varphi\biggr]
- g^{\nu\alpha} \delta_{s} \Gamma_{\alpha\mu}^{\sigma} \partial_{\sigma} \varphi
\nonumber\\
&& +\frac{1}{2} \delta_{\mu}^{\nu} \biggl[ \delta_{s} g^{\alpha\beta} \partial_{\alpha}\varphi \partial_{\beta} \varphi -  2 \delta_{s} g^{\alpha\beta} (  \partial_{\alpha} \partial_{\beta}\varphi - \Gamma_{\alpha\beta}^{\sigma} \partial_{\sigma}\varphi)
+ 2 g^{\alpha\beta} \delta_{s} \Gamma_{\alpha\beta}^{\sigma} \partial_{\sigma} \varphi\biggl]
\nonumber\\
&& - \frac{1}{4} e^{- \varphi} \frac{\delta_{s} g^{\alpha\beta} \partial_{\alpha}\varphi \partial_{\beta} \varphi}{ \sqrt{ (\partial \varphi)^2}} \,
\gamma_{\mu}^{\nu} \,
{\mathcal I}_1 - \frac{1}{2} e^{-\varphi} \sqrt{ (\partial \varphi)^2}\, \delta_{s} \gamma_{\mu}^{\nu} \,{\mathcal I}_1 
- \frac{1}{2} e^{-\varphi} \sqrt{(\partial \varphi)^2}\,  \gamma_{\mu}^{\nu} \,\delta_{s} {\mathcal I}_1 =0,
\label{PERT2}
\end{eqnarray}
where, as in the case of Eq. (\ref{PERT1}), all the quantities not preceded by $\delta_{s}$ in Eq. (\ref{PERT2}) must be understood as evaluated on the background.
Equation (\ref{PERT2}) shall be further simplified by noting that, in the uniform field gauge, the 
scalar fluctuations of $e^{- \overline{\varphi}}$, ${\mathcal I}_1$ and $\gamma_{\mu}^{\nu}$ vanish (i.e. 
$\delta_{s} (e^{- \overline{\varphi}} )= \delta_{s} {\mathcal I}_1 = \delta_{s} \gamma_{\mu}^{\nu} =0$).
Using the decomposition (\ref{S1}) and imposing the uniform dilaton gauge, the $(00)$ and $(0i)$ components of Eq. (\ref{PERT2}) can be respectively written as:
\begin{eqnarray}
&& 3 (\overline{\varphi}' + {\mathcal H}_{s}) \psi' - ( {\overline{\varphi}'}^2 - 3 {\mathcal H}_{s}^2) \phi + 2 \nabla^2 \psi 
- (\overline{\varphi}' + {\mathcal H}_{s}) \nabla^2 E'=0,
\label{00P2}\\
&& (\overline{\varphi}' + {\mathcal H}_{s}) \phi = 2 \psi'.
\label{i0P2}
\end{eqnarray}
Similarly the component $(i\neq j)$  of Eq. (\ref{PERT2}) becomes:
\begin{equation}
\partial_{i}\partial^{j} [ E'' -(\overline{\varphi}' + {\mathcal H}_{s}) E' + \psi - \phi] =0. 
\label{ineqjP2}
\end{equation}
Finally the explicit form of the $(i = j)$ component of Eq. (\ref{PERT2}) is given by:
\begin{eqnarray}
&& 2 \psi'' - 2 ( \overline{\varphi}' + {\mathcal H}_{s}) \psi' - (\overline{\varphi}' + {\mathcal H}_{s}) \phi' - \nabla^2 [ E''  -( \overline{\varphi}' + {\mathcal H}_{s} ) E' + (\psi - \phi)] 
\nonumber\\ 
&&- \phi [ 2 \overline{\varphi}'' - {\overline{\varphi}'}^2 - \frac{a_{s}^2}{2} \frac{\partial V}{\partial \overline{\varphi}} + 
2 {\mathcal H}'_{s} - 5 {\mathcal H}_{s}^2 - 4 {\mathcal H}_{s} \overline{\varphi}'] =0.
\label{ijP2}
\end{eqnarray}
The coefficient of $\phi$ in Eq. (\ref{ijP2}) can be 
expressed as $(a_{s}^2/2) (\partial V/\partial\overline{\varphi}) - V a_{s}^2$ by repeated use of Eqs. (\ref{back1}) and (\ref{back2}).  
In the uniform field gauge the fluctuations of the spatial curvature are given solely in terms of $\psi$:
\begin{equation}
\delta_{s} R^{(3)} \equiv \frac{4}{a_{s}^2}\nabla^2 \psi.
\label{GS88}
\end{equation}
Up to a sign which changes depending on different conventions, the value of  $\psi$  in the uniform field 
gauge coincides with the curvature perturbations customarily indicated by ${\mathcal R}$. 
Strictly speaking ${\mathcal R}$ defined the curvature perturbation on comoving orthogonal 
hypersurfaces. However, since ${\mathcal R}$ is by definition gauge invariant its evolution 
equations can be derived in any gauge, such as the one employed in the present discussion\footnote{In what follows we shall deduce the equation for ${\mathcal R}$ and we shall demonstrate that 
${\mathcal R}$ is not only the correct gauge-invariant variable to be used but it is also frame-invariant.
See, in this respect, also the discussion of appendix \ref{APPA}.  }.

The strategy will therefore to derive the equations first in the string frame by combining  Eqs. (\ref{00P2})--(\ref{i0P2}) and (\ref{ineqjP2})--(\ref{ijP2}).
From Eq. (\ref{i0P2}) we have
$\phi = 2 \psi'/(\overline{\varphi}' + {\mathcal H}_{s})$; inserting this relation into Eq. (\ref{00P2}) the following expression can be readily obtained:
\begin{equation}
\psi' = - \frac{2 ( \overline{\varphi}' + {\mathcal H}_{s})}{(\overline{\varphi}' + 3 {\mathcal H}_{s})^2} \nabla^2\psi 
+ \frac{(\overline{\varphi}' + {\mathcal H}_{s})^2}{(\overline{\varphi}' + 3 {\mathcal H}_{s})^2} \nabla^2 E'.
\label{pspr}
\end{equation}
Equation (\ref{pspr}) holds in the string frame but it can be easily transformed into the Einstein frame by using 
the properties of the uniform dilaton gauge. 
Indeed recalling Eqs. (\ref{rel1}) and (\ref{def1})--(\ref{def2}) we also have that ${\mathcal H}_{s} = {\cal H}_{e} + \varphi'/2$. Since  $\overline{\varphi} = \varphi' - 3 {\mathcal H}_{s}$,  Eq. (\ref{pspr}) can be immediately written as:
\begin{equation}
\psi' = \frac{ 4 {\cal H}_{e}}{{\varphi'}^2} \nabla^2 ( \psi + {\cal H}_{e} E').
\end{equation}
According to Eq. (\ref{GS88}) in the uniform dilaton gauge ${\mathcal R } = - \psi$; in the same gauge 
we can compute the Bardeen potential and the result is $\Psi_{e} = \psi + {\mathcal H}_{e} E'$ (see 
also appendix Eq. (\ref{fin})). We thus obtain from Eq. (\ref{pspr}) the following simple equation ${\mathcal R}' = -  4 ({\mathcal H}_{e}/\varphi^{\prime\,2}) \nabla^2  \Psi_{e}$. Let us finally mention that the relation $\phi = 2 \psi'/(\overline{\varphi}' + {\mathcal H}_{s})$ derived from 
 Eq. (\ref{i0P2}) implies that  Eq. (\ref{ijP2}) is identically satisfied and does not imply further conditions. 
 Indeed inserting $\phi = 2 \psi'/(\overline{\varphi}' + {\mathcal H}_{s})$ into  Eq. (\ref{ijP2}) we have: 
\begin{equation}
\frac{\psi'}{(\overline{\varphi}' + {\mathcal H}_{s})}\biggl[ 2 {\mathcal H}_{s}^{\prime} + 4 {\mathcal H}_{s}^2 - 2 \overline{\varphi}'' + 
a_{s}^2 \frac{\partial V}{\partial \overline{\varphi}}\biggr] =0,
\label{bvan}
\end{equation}
but this is an identity since the expression between square brackets vanishes on the background. In fact, 
inserting the two relations of Eq. (\ref{back1}) into Eq. (\ref{back2}) we obtain that the combination appearing in Eq. (\ref{bvan}) between square brackets 
is bound to vanish.

\subsection{Frame-invariance and scalar sound speed}
To derive the gauge-invariant and frame-invariant evolution of the curvature perturbations 
it is appropriate to rewrite the system of scalar perturbations in a more compact
form by using Eqs. (\ref{back1})--(\ref{back2}):
\begin{eqnarray}
&& E^{\prime\prime} - y_{1} E^{\prime} + \psi - \phi=0,\qquad  y_{1} \phi = 2 \psi^{\prime},
\label{II}\\
&&2 (\psi^{\prime\prime} - y_{1} \psi^{\prime}) = y_{1} \phi^{\prime} +( y_{1}^{\prime} - y_{1}^2) \phi,
\label{III}\\
&& 3 y_{1} \psi^{\prime} - V\,a^2 \phi + 2 \nabla^2 \psi - y_{1} \nabla^2 E^{\prime} =0,
\label{IV}
\end{eqnarray}
where the new background variable $y_{1} = \overline{\varphi}' +  {\mathcal H}_{s}$ has been introduced.
It is now evident from Eqs. (\ref{II}), (\ref{III} and (\ref{IV}) that once the second relation of Eq. (\ref{II}) is inserted into (\ref{III})  an identity 
is swiftly obtained. Using then $y_{1} \phi = 2 \psi^{\prime}$ into Eq. (\ref{IV}) the background equations imply:
\begin{equation}
\psi^{\prime} = - \frac{2 y_{1}}{y_{2}^2 } \nabla^2 \psi + \biggl(\frac{y_{1}}{y_{2}}\biggr)^2 \nabla^2 E^{\prime},
\label{V}
\end{equation}
where the background combination $y_{2} = \overline{\varphi}' + 3 {\mathcal H}_{s} $ has been defined. 
Equation (\ref{V}) has the same dynamical content of Eq. (\ref{pspr}) but it is more practical. 
Let us now take the conformal time derivative of both sides of Eq. (\ref{V}) and replace the 
terms  $\nabla^2 E''$ and $\nabla^2 E'$ by means of  Eqs. (\ref{II}) and (\ref{V}).
The final result of this lengthy but straightforward procedure is the following decoupled equation for $\psi$ :
\begin{equation}
\psi'' - \biggl[ 2 \biggl( \frac{y_{2}}{y_{1}}\biggr) \biggl(\frac{y_{1}}{y_{2}}\biggr)' + y_{1}\biggr] \psi' - 
\biggl(\frac{ y_{1}^2 + 2 y_{1}'}{y_{2}^2}\biggr) \nabla^2 \psi =0.
\label{VI}
\end{equation}
Recalling now the explicit expressions of $y_{1}$ and $y_{2}$ we can rewrite some of the background dependent 
quantities appearing in Eq. (\ref{VI}). In particular the following identity is verified
\begin{equation}
2 \frac{z_{s}^{\prime}}{z_{s}}=  - \biggl[ 2 \biggl( \frac{y_{2}}{y_{1}}\biggr) \biggl(\frac{y_{1}}{y_{2}}\biggr)' + y_{1}\biggr], \qquad 
z_{s} = - 2 \frac{\overline{\varphi}^{\prime} + 3 {\mathcal H}_{s}}{\overline{\varphi}' + {\mathcal H}_{s}} a_{s} e^{-\varphi/2}.
\label{StR0}
\end{equation}
Thus from Eq. (\ref{VI}) the evolution equation for curvature perturbations ${\mathcal R}_{s} = - \psi$ becomes:
\begin{equation}
{\mathcal R}_{s}^{\prime\prime} + 2 \frac{z_{s}^{\prime}}{z_{s}} {\mathcal R}_{s}^{\prime} -  c_{s}^2 \nabla^2 {\mathcal R}_{s} =0,
\label{StR1}
\end{equation}
and the sound speed squared is:
\begin{equation}
c_{s}^2 = \frac{ y_{1}^2 + 2 y_{1}'}{y_{2}^2}=  1+ \frac{2 \overline{\varphi}^{\prime\prime} - 6 {\mathcal H}_{s}^2 + 2 {\mathcal H}_{s} \overline{\varphi}^{\prime}}{\varphi^{\prime\,2}}.
\label{Str1a}
\end{equation}
The second equality in Eq. (\ref{Str1a}) follows directly from the expressions of $y_{1}$ and $y_{2}$; note, in particular, that $y_{2} = \overline{\varphi}^{\prime} + 3 {\mathcal H}_{s} \equiv \varphi^{\prime}$ and $y_{1} = y_{2} - 2 {\mathcal H}_{s}$. It is furthermore easy to prove that a combination of Eqs.  
(\ref{b2in}) and (\ref{b3in}) implies the following identity 
\begin{equation}
6 {\mathcal H}_{s}^2 - 2 \overline{\varphi}^{\prime\prime} + 2 {\mathcal H}_{s} \overline{\varphi}^{\prime} 
+ \frac{\partial V}{\partial\overline{\varphi}} a_{s}^2 =0.
\label{Str1b}
\end{equation}
Inserting Eq. (\ref{Str1b}) into Eq. (\ref{Str1a}) the expression for the sound speed can be easily determined:
\begin{equation}
 \qquad c_{s}^2 = 1 + \frac{\partial V}{\partial \overline{\varphi}} \frac{a_{s}^2}{\varphi^{\prime \,\, 2}}.
\label{StR2}
\end{equation}
The results of Eqs. (\ref{StR1}) and (\ref{StR2}) have a direct counterpart in the Einstein frame where 
\begin{equation}
{\mathcal R}_{e}^{\prime\prime} + 2 \frac{z_{e}^{\prime}}{z_{e}} {\mathcal R}_{e}^{\prime} - c_{e}^2 \nabla^2 {\mathcal R}_{e} =0,
\label{EinR1}
\end{equation}
and 
\begin{equation}
 z_{e}  =  \frac{a_{e} \varphi^{\prime}}{{\mathcal H}_{e}}, \qquad c_{e}^2 = 1 + \frac{\partial V}{\partial \overline{\varphi}}\frac{ e^{\varphi} a_{e}^2 }{\varphi^{\prime\, 2}}.
\label{EinR0}
\end{equation}
When $\varphi \to \varphi$ and ${\mathcal H}_{s} \to {\mathcal H}_{s} = {\mathcal H}_{e} + \varphi^{\prime}/2$ (see also  Eq. (\ref{transition})) we have also have that $z_{s} \to z_{e}$ and $c_{s}^2 \to c_{e}^2$, as expected\footnote{This means, in particular, that once expressed on a given background solution and in terms of the conformal time 
coordinate $c_{s}^2(\tau)\equiv c_{e}^2(\tau)$ and $z^{\prime}_{s}/z_{s} \equiv z^{\prime}_{e}/z_{e}$. }.
  Unlike the Bardeen potentials (which are gauge-invariant but not necessarily 
frame-invariant), the variable ${\mathcal R}$ will then denote the common value of the curvature 
perturbations either in the string or in the Einstein frame (i.e. ${\mathcal R}= {\mathcal R}_{e} = {\mathcal R}_{s}$).
This is in full analogy with the case of the tensor modes of the geometry discussed at the beginning of this section. 

It is natural, at this point, to identify $c_{s}^2$ or $c_{e}^2$ with an effective sound speed. We note that this identification rests on the analogy of Eqs. (\ref{StR1}) and (\ref{EinR1})  with the equation describing the normal modes of a gravitating, irrotational and relativistic fluid firstly discussed by Lukash \cite{lukash}
\begin{equation}
{\mathcal R}^{\prime\prime} + 2 \frac{z^{\prime}}{z} {\mathcal R}^{\prime} - c_{t}^2 \nabla^2 {\mathcal R} =0,\qquad 
 z = a^2 \sqrt{p_{t} + \rho_{t}}/{\mathcal H}, \qquad c_{t}^2 = \frac{p_{t}^{\prime}}{\rho_{t}^{\prime}},
\label{luk}
\end{equation}
where $p_{t}$ and $\rho_{t}$ denote, in the context of Eq. (\ref{luk})
the pressure and the energy density of a perfect and irrotational fluid. 
The canonical normal mode identified in Ref. \cite{lukash}  is invariant under infinitesimal coordinate transformations as required in the context of the  Bardeen formalism \cite{bard1} (see also \cite{lif}). The subsequent analyses of Refs.  \cite{bard2} follow the same logic of \cite{lukash} but in the case of scalar field matter;  the normal modes of Refs. \cite{lukash,bard2}  coincide with the (rescaled) curvature perturbations on comoving orthogonal hypersurfaces \cite{bard1,bard2}.  Owing to the analogy of Eq. (\ref{EinR1}) with the Lukash equation (\ref{luk}) it would be 
tempting to carry this analogy even further by recalling that the evolution equations in the 
Einstein frame can be phrased in terms of an effective energy density 
$\rho_{e}$ supplemented by an effective pressure (see, in this respect, Eq. (\ref{pr}) and discussion therein). 
This literal correspondence is however misleading 
since  $z_{e} \neq a_{e}^2 \sqrt{p_{e} +\rho_{e}}/{\mathcal H}_{e}$ 
as it would follow from a naive combination of  Eqs. (\ref{pr}) and (\ref{luk}). 
This is why, in the present context, we shall always qualify the sound speed as effective.

The evolution equations of the linearized fluctuations discussed here all local in time however it is true 
that the presence of the potential induces a sound speed for the scalar modes of the geometry.  This 
 is particularly clear if we notice that the sound speed goes to 1 in the limit 
 of vanishing nonlocal potential: if the potential would just be local (i.e. only dependent on $\varphi$) the 
 sound speed would coincide with the speed of light. It is therefore correct to conclude
  that the presence of a potential depending on the shifted dilaton $\overline{\varphi}$ induces a scalar sound speed.

\subsection{Imaginary sound speed and gradient instability}
We are now going to show that a four-dimensional curvature bounce connecting 
analytically two duality related solutions leads to a gradient instability. For those wavenumbers $k$ exceeding the typical 
scale of the bounce (of the order of $1/t_{0}$ in the example discussed below), 
the solutions of the evolution equation for ${\mathcal R}$ instead 
of oscillating are exponentially amplified.  To demonstrate this point let us focus on the following 
well known solution (see \cite{nonloc1}):
\begin{eqnarray}
&& V (\overline{\varphi}) = - V_0 e^{ 4\overline{\varphi} },\qquad H_{s} = \frac{1}{\sqrt{3} \, t_{0} \, \sqrt{(t/t_0)^2 +1}},
\label{pot1}\\
&& \overline{\varphi} = \varphi - 3 \ln a_s(t) =- \frac{1}{2} \log{ [ 1 + (t/t_{0})^2]} + \varphi_{0},
\label{dilaton1}
\end{eqnarray}
where $t$ is the cosmic time coordinate. Equations (\ref{pot1}) and (\ref{dilaton1}) are a solution of the background equations in the string frame. In particular 
to solve Eqs. (\ref{b1}) and (\ref{b2}) the constants $t_{0}$, $V_{0}$ and $\varphi_{0}$ appearing Eqs. (\ref{pot1})--(\ref{dilaton1}) must satisfy $t_0\,e ^{2 \varphi_0}\,\sqrt{V_{0}} =1$. The potential (\ref{pot1}) is always negative and, in practice, it only modifies the solution for $|t| < {\mathcal O}(t_{0})$.  Conversely in the 
asymptotic region the potential is always negligible in comparison with the remaining terms of the equations (i.e. $V \ll 3 H_{s}^2$ and $V \ll \dot{\overline{\varphi}}^2$);
thus for $|t| > {\mathcal O}(t_{0})$ the approximate solutions are $a_{s} \simeq (- t/t_{0})^{-1/\sqrt{3}}$ (for $t \ll - t_{0}$) and $a_{s} \simeq ( t/t_{0})^{1/\sqrt{3}}$ (for $t \gg t_{0}$). These asymptotic solutions can be derived from Eqs. (\ref{b1}) and (\ref{b2}) by neglecting the potential altogether.

If we now compare Eqs. (\ref{T1}) and (\ref{StR1}) in the light of the solution (\ref{pot1}) and (\ref{dilaton1}) 
we conclude that the evolution equation for the tensor modes obtained in Eq. (\ref{T1}) does not lead to any gradient instability.
The effective sound speed $c_{s}$ appearing in Eq. (\ref{StR1}) can instead become imaginary 
when $|t| < t_{0}$, i.e. exactly in the regime where the potential modifies the asymptotic ``vacuum" solutions: 
this means that in the limit $k t_{0} > 1$ and for $|t| < t_{0}$ the corresponding Fourier modes of ${\mathcal R}$ are
exponentially amplified. To prove this statement we can compute $c_{s}^2$ and simply demonstrate that it is not positive semidefinite. Indeed 
using Eqs. (\ref{pot1}) and (\ref{dilaton1}) inside Eq. (\ref{StR2}) we obtain
\begin{eqnarray}
c_s^2(t) &=&1 +  \frac{1}{\dot{\varphi}^2}\biggl(\frac{\partial V}{\partial \overline{\varphi}}\biggr)
\nonumber\\
&=&  1 + \frac{4}{-3 - 4\,(t/t_{0})^2 + 2\,{\sqrt{3}}\,(t/t_{0})\,{\sqrt{1 + (t/t_{0})^2}}},
\label{ex}
\end{eqnarray}
where the first equality merely follows from Eq. (\ref{StR2}) by recalling the definition of the cosmic time parametrization (i.e. $a_{s}(\tau) d\tau = d t$).
According to  Eq. (\ref{ex})  we have that $c_{s}^2(t)\to 1$ away from the bounce (i.e. for $| t| > {\mathcal O}(t_{0})$) while\footnote{As an example, for two particular values of $t/t_{0}$,  we have $c_{s}^2(0) = -1/3$ and $c_{s}^2(t_{0}/\sqrt{2}) = -1$. }
 $c_{s}^2\to -1$ for $|t| \leq t_{0}$. For $|t|< t_{0}$ the square of the effective sound speed can be expanded as
 \begin{equation}
c_{s}^2(t) = -\frac{1}{3} -\frac{8}{3 \sqrt{3}} \biggl(\frac{t}{t_{0}}\biggr) + {\mathcal O}(t^2/t_{0}^2).
\label{ex2}
\end{equation}
When $k t_{0} \ll 1$ (or for $k t_{0} \sim {\mathcal O}(1)$) values $c_{s}^2 < 0$ around the origin are never 
problematic and typically lead to a steeply increasing spectrum of scalar modes \cite{nonloc1}. 
However when  $k t_{0} \gg 1$ we can have that  the Laplacian of Eq. (\ref{StR1}) 
becomes $- c_{s}^2(t)\nabla^2{\mathcal R}\to k^2 c_{s}^2(t)$; but since this term gets sharply negative all modes $k t_{0}$ 
are exponentially amplified for $|t| <t_{0}$. Since the whole description 
of curvature perturbations is frame-invariant, the occurrence of the gradient instability is also frame-invariant 

If we would assume that gradient instabilities are generically present in bouncing model the problem could be cured (or at least alleviated) by the arbitrary addition of further terms 
in the action of the scalar modes of perturbations.  For instance the evolution equations 
of curvature perturbations could be written as:
\begin{equation}
S^{(2)} = \frac{1}{2} \int d^{3} x \int d \tau z_{s}^2 \,\biggl[ (\partial_{\tau} {\mathcal R})^{2} - 
c_{s}^2 \delta^{ij} \partial_{i} {\mathcal R}\partial_{j} {\mathcal R}
+ \frac{q_{1}(\tau)}{M^2} (\nabla^2 {\mathcal R})^2 + \frac{q_{2}(\tau)}{M^4} \delta^{ij} (\partial_{i} \nabla^2 {\mathcal R}) (\partial_{j} \nabla^2 {\mathcal R}) \biggr], 
\label{cur1}
\end{equation}
where $M$ is a mass scale possibly related to the maximal curvature scale reached at the bounce, i.e. 
$M = {\mathcal O}(H_{max} a_{max})= {\mathcal O}( 1/\tau_{max})$. In the limit $q_{1}(\tau) = q_{2}(\tau) =0$ 
the variation of the second-order action of Eq. (\ref{cur1}) leads immediately to Eq. (\ref{StR1}). 
Conversely if these two terms 
are present the evolution equation of ${\mathcal R}$ can be written as:
\begin{equation}
\partial_{\tau}[ z_{s}^2 \partial_{\tau} {\mathcal R}] - c_{s}^2 \, z_{s}^2 \nabla^2 {\mathcal R} - \frac{q_{1}}{M^2} z_{s}^2  \nabla^4 {\mathcal R}- \frac{q_{2}}{M^4} z_{s}^2 \nabla^{6} {\mathcal R} =0.
\label{cur2}
\end{equation}
Let us now suppose, for the sake of concreteness, that the two background dependent functions $q_{1}(\tau)$ and $q_{2}(\tau)$ 
will be continuous and differentiable everywhere. Going to Fourier space and rearranging the terms of Eq. (\ref{cur2}) 
\begin{equation}
 {\mathcal R}_{k}^{\prime\prime} + 2 \frac{z_{s}^{\prime}}{z_{s}} {\mathcal R}_{k}^{\prime} + k^2 c_{tot}^2 {\mathcal R}_{k}=0,\qquad
c_{tot}^2 = c_{s}^2 - q_{1}\kappa^2 + q_{2}\kappa^4,
\label{cur3}
\end{equation}
where $\kappa= k/(a_{max} H_{max})$. Even if  $c_{s}^2$ becomes negative in Eq. (\ref{cur3}), $c_{tot}^2$ may well be positive 
depending on the sign of $q_{1}$ and $q_{2}$. More specifically let us suppose that $c_{s}^2$ becomes negative 
in the neighbourhood of the origin (say for $|\tau | <\tau_{0}$); even if, in the same region,  $q_{1}(\tau) > 0$ we will have that 
$c_{tot}^2 \geq 0$ for $\kappa \gg 1$ provided $q_{2}(\tau) \geq 0$ when  $|\tau | <\tau_{0}$. 
Note that corrections such as the ones appearing in Eq. (\ref{cur1}) arise naturally 
when applying the methods of effective field to the analysis of generic theories of inflation with a single inflaton field \cite{wein}.
In generic theories of inflation the dependence of the action on 
the inflaton field is unconstrained and a similar analysis translates to the case of bouncing models even 
if a shift symmetry may be imposed on the action so that the Lagrangian density will involve 
only spacetime derivatives rather than the field itself \cite{GI1,GI2}. 
We will now argue that the gradient instability of Eqs. (\ref{ex}) and (\ref{ex2}) is a property of the solution but not necessarily a property of the model. In other words we can expect that by changing 
the solution also the gradient instability might disappear. 

\renewcommand{\theequation}{4.\arabic{equation}}
\setcounter{equation}{0}
\section{Taming the gradient instability}
\label{sec4}
The gradient instability arising in the bouncing models based on the action (\ref{action1}) and scrutinized 
in section \ref{sec3} is nongeneric. For this purpose, a purely qualitative analysis shows that that the sign of the effective scalar 
sound speed, in both frames, is determined by the sign of the derivative of 
$V(\overline{\varphi})$ with respect to $\overline{\varphi}$.
Let us now consider the form of the potential reported in Eq. (\ref{pot1}): away from the 
bounce, the potential and its derivative are both subleading and this is the reason why, incidentally,
the solution in the asymptotic regions exactly matches the (duality related) vacuum solutions. 
Near the bounce the contribution 
of the potential and of its derivative always enters the effective scalar sound speed 
with a negative sign: this is ultimately the reason why $c_{s}^2<0$  for $|t| < t_{0}$.
To construct examples where the gradient instability is tamed 
we must therefore consider more seriously those models where 
$\partial V/\partial\overline{\varphi}$ does not have a definite sign in the bouncing region.

\subsection{A class of semi-realistic backgrounds}
Guided by the logic spelled out in the previous paragraph let us therefore consider, for the sake of concreteness, the following class of potentials
\begin{equation}
V(\overline{\varphi}) = \frac{V_{1}}{\cosh^2{[\beta (\overline{\varphi} - \overline{\varphi}_{1})]}},
\label{pot2}
\end{equation}
whose derivative with respect to $\overline{\varphi}$ changes sign for $\overline{\varphi} = \overline{\varphi}_{1}$.
Inserting Eq. (\ref{pot2}) into Eqs. (\ref{Eone}), (\ref{Etwo}) and (\ref{Ethree}) the evolution equations of the background can be 
solved in explicit terms. Without loss of generality we shall focus on the solution 
in the Einstein frame and in the conformal time parametrization which is 
the most convenient for the analysis of the inhomogeneities. 
Note, in particular, that in the Einstein frame the following useful relation can be obtained 
after repeated combinations of Eqs. (\ref{Eone}), (\ref{Etwo}) and (\ref{Ethree}):
\begin{equation}
\frac{\partial}{\partial \tau} (\varphi^{\prime} + 2 {\mathcal H}_{e}) + 2 {\mathcal H}_{e} (\varphi^{\prime} + 2 {\mathcal H}_{e}) =0.
\label{pot2a}
\end{equation}
The bouncing solution corresponding to Eq. (\ref{pot2}) can then be written as: 
\begin{eqnarray}
{\mathcal H}_{e}(\tau) &=& \frac{ {\mathcal H}_{1}}{2 \beta \sqrt{x^2 + 1}},\qquad x = \frac{\tau}{\tau_{1}}
\label{Eeighta}\\
 \varphi(\tau) &=& \overline{\varphi}_{1} - \frac{1}{\beta} \log{[x+ \sqrt{x^2 +1}]},
 \label{Eeightb}\\
 a(\tau) &=& a_{1} \biggl[ x+ \sqrt{x^2 +1}\biggr]^{\frac{1}{2\beta}},
 \label{Eeightc}
\end{eqnarray}
where $\tau_{1}$ denotes the typical scale of the bounce and ${\mathcal H} = a_{1} H_{1} = 1/\tau_{1}$; in the rescaled coordinate $x$ the bouncing 
region corresponds to $|x| < 1$. To satisfy consistently all the equation the relation between the integration constants 
${\mathcal H}_{1}$, $V_{1}$ and $\overline{\varphi}_{1}$ 
must be given by ${\mathcal H}_{1} = \beta \sqrt{V_{1}} \exp{[\overline{\varphi}_{1}/2]}$. The solution (\ref{Eeighta}), (\ref{Eeightb}) and (\ref{Eeightc})
holds in the Einstein frame and (most importantly) in the conformal time parametrization\footnote{This new solution superficially 
resembles the one of Eqs. (\ref{pot1}) and (\ref{dilaton1}): it should be however clear that the two solutions are totally different 
since Eqs. (\ref{pot1}) and (\ref{dilaton1}) hold in the string frame and their Einstein frame form cannot be obtained analytically 
for the whole time range but only in the asymptotic regions. Furthermore, as stressed, the properties of the potential are completely different.}.

From Eqs. (\ref{Eeighta}) and (\ref{Eeightb}) the total sound speed of the scalar fluctuations can be easily computed from the general expression and 
the result of this manipulation is given by:
\begin{equation}
c_{e}^2(\tau) = 1 + 2 \beta \frac{x}{\sqrt{x^2 +1}}.
\label{SS1}
\end{equation}
Note that once expressed in the conformal time coordinate and on the explicit solution 
the sound speeds in the string and Einstein frames coincide (i.e. $c_{e}^2(\tau) = c_{s}^2(\tau)$).
From Eq. (\ref{SS1}) the overall sign of $c_{e}^2$ is determined by the value of beta. In principle 
we could concoct a value of beta leading to $c_{s}^2 >0$ and simultaneously describing an accelerated 
contraction before the bounce. However, for the sake of concreteness, it seems more useful 
to compute the spectrum and select those values of $\beta$ allowing for a quasi-flat 
spectrum of curvature perturbations, as we shall show in a moment.

Even if Eq. (\ref{Eeightc}) does not allow for an analytic connection between the conformal and the 
cosmic time coordinate over the whole range of variation of $\tau$ (or $x$) the relation of $\tau$ to $t$ 
can easily be determined piecewise in the asymptotic regions.
In particular in the limits $\tau \ll -\tau_{1}$ and $\tau \gg \tau_{1}$ the scale factor evolves, respectively, as 
\begin{eqnarray}
\lim_{\tau \ll - \tau_{1}} a_{e}(\tau) \to \biggl(-\frac{\tau}{\tau_{1}}\biggr)^{- \frac{1}{2\beta}} = \biggl(- \frac{t_{e}}{t_{1}}\biggr)^{\frac{1}{1 - 2 \beta}},
\label{SS1a}\\
\lim_{\tau \gg \tau_{1}} a_{e}(\tau) \to \biggl(\frac{\tau}{\tau_{1}}\biggr)^{ \frac{1}{2\beta}} = \biggl( \frac{t_{e}}{t_{1}}\biggr)^{\frac{1}{1 + 2 \beta}},
\label{SS1b} 
\end{eqnarray}
where the last two equalities at the right hand side follow from the relation between the cosmic and the conformal time coordinate (i.e. $a_{e}(\tau) \, d\tau = dt_{e}$). The solution of Eqs. (\ref{pot2}) and (\ref{Eeighta})--(\ref{Eeightb}) is a special case of a more general set of solutions 
characterized by 
\begin{equation}
\varphi= \overline{\varphi}_{1} - 2 \ln{a_{e}} + f_{1} \int_{\tau_{i}}^{\tau} \biggl[\frac{a_{1}}{a_{e}(\tau^{\prime})}\biggr]^2 \, d\tau^{\prime},
\label{SS2}
\end{equation}
where $f_{1}$ is a dimensional constant. By making use of Eq. (\ref{SS2}) we can easily write the explicit form of the sound speed.  More specifically 
thanks to  Eqs. (\ref{Eone}), (\ref{Etwo}) and (\ref{Ethree}) the result can be expressed as $c_{s}^2 = 1 - {\mathcal N}_{s}/{\mathcal D}_{s}$ where 
${\mathcal N}_{s}$ and ${\mathcal D}_{s}$ can be written, respectively, as:
\begin{eqnarray}
{\mathcal N}_{s}(x) &=& 16 \overline{{\mathcal H}}_{e} \frac{\partial \overline{{\mathcal H}}_{e} }{\partial x} + 4 \biggl(\frac{a_{1}}{a_{e}}\biggr)^2 \lambda \biggl( \frac{\partial \overline{{\mathcal H}}_{e} }{\partial x} - 10 \overline{{\mathcal H}}_{e}^2\biggr) + \lambda^3\biggl(\frac{a_{1}}{a_{e}}\biggr)^{6},
\nonumber\\
{\mathcal D}_{s}(x) &=& \biggl[\lambda \biggl(\frac{a_{1}}{a_{e}}\biggr)^2 - 2 \overline{{\mathcal H}}_{e} \biggr]^2 \biggl[\lambda \biggl(\frac{a_{1}}{a_{e}}\biggr)^2 +4\overline{{\mathcal H}}_{e} \biggr].
\label{SS3}
\end{eqnarray}
Note that $\overline{{\mathcal H}}_{e}$ is dimensionless, i.e.  ${\mathcal H}_{e} = {\mathcal H}_{1} \overline{{\mathcal H}}_{e}$ and 
$\lambda = f_{1}/{\mathcal H}_{1}$ is also a dimensionless constant. For $a_{e} \gg a_{1}$ (i.e. away from the bounce) 
the terms weighted by $\lambda$ are always negligible and Eq. (\ref{SS3}) reduces to 
\begin{equation}
c_{e}^2(\tau) = 1 - \frac{1}{\overline{\mathcal H}_{e}} \frac{\partial \overline{\mathcal H}_{e}}{\partial x}.
\label{SS4}
\end{equation}

As far as the power spectra are concerned what matters is not the behaviour in the region 
$|x| < 1$ (i.e.  $|\tau| < \tau_{1}$) but rather the evolution in the asymptotic regions and, in 
particular, in the pre-bounce stage. In this respect we can mention that there are 
different solutions with the same asymptotic behaviour of Eq. (\ref{Eeighta}) 
which have however a different analytical structure in the region $|x|<1$.
A class of models sharing this property is given by 
\begin{equation}
{\mathcal H}_{e}  = \frac{\alpha {\mathcal H}_{1} }{ (x^{2 \gamma} +1)^{1/(2 \gamma)}}, \qquad c_{e}^2 = 1 + \frac{1}{\alpha} \frac{x^{2 \gamma -1}}{(1 + x^{2\gamma})^{1 - \frac{1}{2\gamma}}}.
\label{furth}
\end{equation}
where $\gamma=1,\,2,\, .\,.\,.\,$ is an integer. Note that the solution of Eq. (\ref{Eeightc}) corresponds to $\gamma = 1$ in Eq. (\ref{furth}). 
For $\gamma > 1$ the analytical behaviour of the solution is different from the ones discussed before. The relevant point, however, 
 is not the specific analytic form of the solution but rather the positivity of 
$c_{e}^2$ which can be realized in different ways.  

\subsection{The spectrum of inhomogeneities}

The results obtained so far demonstrate that the occurrence of the gradient instability is 
nongeneric when the bounce is regularized by means of a nonlocal dilaton potential. We now 
want to fix the value of $\beta$ in a more realistic realistic way. In what follows 
we shall analyze specifically the class of models of Eqs. (\ref{SS1}) and (\ref{SS1a}) and show that
the spectrum of curvature perturbations may indeed be flat when $\beta \simeq -1/4$. 
From Eq. (\ref{SS1a}) the first and second (cosmic) time derivatives of the scale factor in the asymptotic region 
$t < - t_{1}$ are given, respectively, by:
\begin{equation}
\dot{a}_{e} = -\frac{1}{(1 - 2\beta) t_{1}} \biggl(-\frac{t_{e}}{t_{1}}\biggr)^{\frac{2\beta}{ 1 - 2 \beta}}, \qquad 
\ddot{a}_{e} = \frac{2 \beta}{t_{1}^2 \,(1 - 2 \beta)^2} \biggl( - \frac{t_{e}}{t_{1}}\biggr)^{\frac{4 \beta -1}{1 - 2 \beta}}.
\label{SS1c}
\end{equation}
Equation (\ref{SS1c}) indicates that the case of accelerated contraction (i.e. $\dot{a}_{e} < 0$ and 
$\ddot{a}_{e} <0$) is realized when $\beta <0$. Thus values  $\beta<0$  will be regarded as the most physical ones.
Different choices are possible but they are not central to the present discussion. 

The amplified curvature perturbations in the  case where the relevant modes exited the Hubble radius for $\tau < - \tau_{1}$ will now be computed. 
This terminology is inaccurate but often used. What matters here is not the Hubble radius itself but the nature of the 
pump field governing the evolution of the scalar modes. The  {\em exit}  refers here to the moment where pump field $z_{e}^{\prime\prime}/z_{e}$ 
equals approximately $c_{e}^2 k^2$ (see below Eq. (\ref{fk})). At this turning point the solution of the mode functions change behaviour. 
Since the initial conditions of the curvature perturbations are set by quantum mechanics, it is essential to remind on the canonical structure of the problem.
To be specific we could say that the quantization of the fluctuations follows exactly 
the same steps outlined by Lukash \cite{lukash} when discussing the 
scalar modes of an irrotational relativistic fluid: unlike Ref. \cite{lukash}  the sound speed 
is only effective. The evolution equation (\ref{EinR1}) can be obtained by functional variation from the following 
action 
\begin{equation}
 S_{{\mathcal R}} = \frac{1}{2} \int d^4 x\,\, z_{e}^2 \biggl[{ {\mathcal R}'}^2 - c_{e}^2 (\partial_{i} {\mathcal R})^2\biggr].
\label{Raction1}
\end{equation}
which falls into the same equivalence class of Eq. (\ref{cur1}).  The normal modes of Eq. (\ref{Raction1}) are $ q = z_{e} {\mathcal R}$; inserting the 
normal modes in Eq. (\ref{Raction1}) and dropping an irrelevant total time derivative we get: 
\begin{equation}
S_{q}= \frac{1}{2} \int d^4 x \biggl[ {q'}^2 - c_{e}^2(\partial_{i} q)^2 +\frac{z''}{z} q^2\biggr],
\label{qaction}
\end{equation}
so that a convenient form of the canonical Hamiltonian can be obtained
\begin{equation}
H(\tau)= \frac{1}{2} \int d^{3} x \biggl[ \tilde{\pi}_{q}^2 +  (\partial_{i} q)^2 - \frac{z''}{z} q^2\biggr].
\label{hamscal3}
\end{equation}
where $\tilde{\pi} = q'$. We can therefore promote $q$ and $\pi$ to field operators obeying equal-time 
commutation relations, i.e. $[\hat{q}(\vec{x},\tau), \hat{\pi}(\vec{y},\tau)] = i \delta^{(3)}(\vec{x} - \vec{y})$ 
in units $\hbar=1$. We eventually want to compute the spectrum of $\hat{{\mathcal R}}= \hat{q}/z_{e}$ so 
that we can write the Fourier representation directly for $\hat{{\mathcal R}}$:
\begin{equation}
\hat{{\mathcal R}}(\vec{x}, \tau) = \frac{1}{(2\pi)^{3/2}} \int d^{3} k \biggl[ F_{k}(\tau) \hat{a}_{\vec{k}}e^{- i \vec{k}\cdot\vec{x}} + 
 F_{k}^{*}(\tau) \hat{a}_{\vec{k}}^{\dagger} e^{ i \vec{k}\cdot\vec{x}} \biggr],
 \label{SS1ca}
\end{equation}
where $[\hat{a}_{\vec{k}}, \hat{a}^{\dagger}_{\vec{p}}] = \delta^{(3)}(\vec{k} - \vec{p})$.
The evolution of the mode functions $F_{k}(\tau)$ and $F_{k}^{*}(\tau)$  can be immediately deduced from 
Eqs. (\ref{EinR1}), (\ref{Raction1}) and (\ref{qaction}). More specifically $F_{k}$ will obey 
\begin{equation}
F_{k}^{\prime\prime} + 2 \frac{z_{e}^{\prime}}{z_{e}} F_{k}^{\prime} + k^2 c_{e}^2 F_{k} =0.
\label{Fk}
\end{equation}
The mode function $f_{k} = z_{e} F_{k}$ will instead follow the same equation obeyed by $\hat{q}$ in the Heisenberg representation: 
\begin{equation}
f_{k}^{\prime\prime} + \biggl[ k^2 c_{e}^2 - \frac{z_{e}^{\prime\prime}}{z_{e}} \biggr] f_{k}=0.
\label{fk}
\end{equation} 
 Using the rescaled conformal time coordinate Eq. (\ref{fk}) becomes\footnote{The rescaled variable $x = \tau/\tau_{1}$ appearing ubiquitously in this section should not be confused 
 with the spatial coordinate $\vec{x}$. }:
\begin{equation}
 \frac{d^2 f_{k}}{ d x^2 }  + \biggl[ \kappa^2 c_{e}^2 - \frac{1}{z_{e}} \frac{d^{2} z_{e}}{d x^2}\biggr] f_{k} =0,\qquad f_{k}(\tau) = z_{e}(\tau) F_{k}(\tau), 
\label{SS1cb}
\end{equation}
where $c_{e}^2(x)$ has been already written in Eq. (\ref{SS1}) while the second term appearing inside the 
 squared brackets of Eq. (\ref{SS1cb}) is given by:
\begin{equation}
\frac{1}{z_{e}} \frac{d^{2} z_{e}}{d x^2} =  \frac{(x^2 + 1) - 2 \beta x \sqrt{x^2+1}}{4 \beta^2 ( x^2 + 1 )^2} \to \frac{1}{2\beta}\biggl(\frac{1}{2\beta} +1\biggr)\frac{1}{x^2}.
\label{SS1d}
\end{equation} 
The limit in Eq. (\ref{SS1d}) follows easily from the first term when $x < -1$ (i.e. $\tau < -\tau_{1}$).  
Consequently the solution of Eq. (\ref{SS1d}) with the correct boundary conditions can be written, for $\beta < 0$, as:
\begin{equation}
F_{k}(\tau) = \frac{1}{ z_{e}(\tau)\sqrt{2\,k\,c_{e}}} {\mathcal N} \sqrt{ - k\,c_{e} \tau} H_{\mu}^{(1)}(- c_{e} k \tau), \qquad \mu = \biggl|\frac{1}{2\beta} + \frac{1}{2} \biggr|, 
\label{SS1e}
\end{equation}
where ${\mathcal N}=  e^{i \alpha/2} \,\sqrt{\pi/2} $ and $\alpha= i(\mu + 1/2)\pi$. The two-point function computed from Eq. (\ref{SS1ca}) is then given by
\begin{equation}
\langle \hat{{\mathcal R}}(\vec{x},\tau) \hat{{\mathcal R}}(\vec{x}+\vec{r},\tau)\rangle= \int d\ln{k}\, {\mathcal P}_{{\mathcal R}}(k,\tau)\, \frac{\sin{k r}}{k r},\qquad {\mathcal P}_{{\mathcal R}}(k,\tau) =\frac{k^3}{2\pi^2} |F_{k}(\tau)|^2.
\label{SS1f}
\end{equation}
The expectation value in Eq. (\ref{SS1f}) is performed over the state minimizing 
the Hamiltonian of the fluctuations.  Since in the limit $\tau \to - \infty$ the space-time 
is flat the initial state is well defined.
The scale-invariant limit for the large-scale modes is therefore realized when $ \mu \simeq 3/2$ which implies that 
for $\beta \simeq -1/4$ the spectrum of curvature perturbations is quasi-flat and the  scale-invariant limit 
is reached in the case $\beta\to -1/4$. In summary we can say that the gradient instability 
does not generically appear in bouncing cosmologies constructed from the action (\ref{action1}). 
We even presented a series of examples where the bounce is correctly regularized, the gradient instability does not arise 
and the Universe evolves from a stage of decelerated contraction.

\subsection{Post-bounce evolution and potential drawbacks} 
The sound speed obtained within the strategy presented in this paper does not lead to a gradient 
instability (i.e. $c_{e}^2$ is correctly positive semidefinite) but the condition 
of subluminal sound speed (i.e. $c_{e}^2 \leq 1$) is not always respected. This is 
an improvement in comparison with the original instability. It is however clear that 
this class of models can only be viewed as semi-realistic even if there are other bouncing 
models with similar drawbacks \cite{GI1,GI2}. Another point of concern is that the obtained bouncing backgrounds do not exit to radiation but to an expanding solution. As argued in the past this problem is closely related 
to the dilaton stabilization. This drawback is common to other scenarios 
and it is potentially very serious. In this respect the idea that backreaction effects can
 produce dynamically a radiation-dominated background  
has been previously discussed. The possibility of a  gravitational reheating of the Universe was  pointed out by various 
authors \cite{BR} and here we shall follow, in particular, the approaches developed by Parker and 
Ford. In \cite{nonloc1} a model of gravitational heating of the 
cold bounce has been proposed by considering  the effects coming from the 
production of Abelian gauge bosons which are directly coupled to $\varphi$ and are copiously produced. 
The frequencies that are maximally amplified are comparable with the typical curvature of the Universe at the bounce
and effectively behave like a gas of massless gauge bosons. Their 
energy density for $ \tau > \tau_{1}$ will be\footnote{See Eq. (4.11) in the second paper of Ref. \cite{nonloc1} and discussion therein.}
$\rho_{r}(\tau) = \epsilon_{0} H_{1}^4 (a_{1}/a)^4$  with $\epsilon_{0} \simeq 0.2$
While fields of different spin will contribute with similar values of $\epsilon_{0}$, in the post-bounce 
regime the dynamically produced radiation may quickly dominate and stabilize the evolution 
of $\varphi$ as argued in the context of specific solutions (see, in particular, the first paper of Ref. \cite{nonloc1}).

\renewcommand{\theequation}{5.\arabic{equation}}
\setcounter{equation}{0}
\section{Concluding remarks}
\label{sec5}
Bouncing models may experience gradient instabilities when
the effective sound speed of the fluctuations becomes imaginary 
for some of the time range where the background solution is defined. 
While this pathology can happen either in the case case of the 
tensor or in the case of the scalar modes, the cures vary depending 
on the specific context. 

Gradient instabilities arise in a class of bouncing models regularized by the 
 presence of a nonlocal dilaton potential which is a scalar under general coordinate transformations but 
depends on the values of the dilaton at two different space-time 
points. Even if the potential is nonlocal in field 
space the evolution equations of the background and of the 
fluctuations are perfectly local in time.  The potential 
is invariant under $T$-duality transformations and may also  
arise in cosmological models inspired by double field theory.
If the dilaton potential is only relevant 
around the bounce and negligible elsewhere the asymptotic background 
solutions follow from the corresponding equations 
with vanishing potential. In this case the
effective sound speed of the scalar modes becomes imaginary for typical time scales 
comparable with the size of the bounce.
The tensor modes are automatically gauge-invariant and frame-invariant and 
do not suffer any instability. 

Following a simple heuristic criterion dictated by the properties of the scalar sound speed, 
different classes of solutions can be found where 
the instability does not arise and the effective sound speed is never imaginary.
The gradient instabilities are tamed for a set of solutions leading, 
incidentally, to a quasi-flat spectrum of curvature inhomogeneities. 
We  suggest or speculate that the exit to radiation (a long standing problem 
of bouncing scenarios) can be naturally addressed by considering 
the backreaction effects on non-conformally coupled species 
as suggested long ago by Ford and Parker. The current proliferation of bouncing scenarios suggests
a number of interesting possibilities which will certainly mature in the years to come.
According to some, realistic scenarios competitive with inflation, are already in sight.
The viewpoint of this paper is more modest and we just regard the present
model as a useful but not yet ultimate theoretical laboratory. 
\newpage 

\begin{appendix}

\renewcommand{\theequation}{A.\arabic{equation}}
\setcounter{equation}{0}
\section{Gauge-invariance and frame-invariance}
\label{APPA}
A series of technical results that have been employed in the bulk of the paper will 
now be presented and derived in a self-contained perspective.
We shall focus on the case of four space-time 
dimensions even if most of the obtained results can easily be 
generalized to higher dimensions \cite{nonloc1,nonloc2}.
The relation between the metric tensors in the Einstein and in the string 
frames\footnote{As in the bulk of the paper the subscripts $e$ and $s$ will distinguish the 
quantities evaluated, respectively, in the Einstein and in the string frames.} is given by:
\begin{equation}
g_{\mu\nu}^{(e)}= e^{-\varphi} g_{\mu\nu}^{(s)}, \qquad \varphi_{s} = \varphi_{e} = \varphi.
\label{RR1}
\end{equation}
The metric fluctuations admit a frame-invariant 
and gauge-invariant description.
In the uniform field gauge \cite{hw1,hw2} the scalar perturbation 
variables are unaltered under a frame redefinition and this property 
makes this choice particularly effective. Conversely the tensor modes of the 
geometry are automatically frame-invariant and also gauge-invariant.
The above statements will now be scrutinized in some detail.

\subsection{Frame-invariance of the tensor modes}
By perturbing Eq. (\ref{RR1}) it is immediately clear that the
tensor modes of the geometry are unaltered when going
from one frame to the other. Let us consider the tensor fluctuation of Eq. (\ref{RR1}):
\begin{equation}
\delta_{t} g_{\mu\nu}^{(e)} = e^{- \varphi} \,\,\delta_{t} g_{\mu\nu}^{(s)},
\label{TT}
\end{equation}
where, as in Eq. (\ref{T0}), $\delta_{t}$ denotes  the tensor fluctuation of the corresponding quantity.
The fluctuation of $\varphi$ (i.e. $\delta_{s} \varphi = \chi$) affects the scalar but not the 
tensor modes of the geometry. Therefore recalling Eq. (\ref{T0}) and its Einstein frame 
analog we have, from Eq. (\ref{TT}), that $a_{e}^2\,  h_{ij}^{(e)} = e^{-\varphi} 
a_{s}^2 h_{ij}^{(s)}$. Since the scale factor (and the extrinsic curvature)
transform as\footnote{We consider here 
the case of conformally flat background geometries in each frame i.e. $g^{(e)}_{\mu\nu} = a_{e}^2(\tau) \eta_{\mu\nu}$ 
and $g^{(s)}_{\mu\nu} = a_{s}^2(\tau) \eta_{\mu\nu}$.}
\begin{equation}
a_{s} = e^{\varphi/2} a_{e},\qquad {\cal H}_{s} = {\cal H}_{e} + \frac{\varphi'}{2},
\label{transition}
\end{equation}
we obtain from Eq. (\ref{TT}) that $h_{ij}^{(e)} = h_{ij}^{(s)} = h_{ij}$ as anticipated 
in section \ref{sec3}.

\subsection{Frame-invariant variables for the scalar modes}
The scalar fluctuations 
of the geometry in the two frames are in principle different and they are related as:
\begin{equation}
\delta_{s} g_{\mu\nu}^{(e)} = e^{-\varphi}[ - \chi \overline{g}_{\mu\nu}^{(s)} + 
\delta_{s} g_{\mu\nu}^{(s)}], \qquad \chi_{s} = \chi_{e} = \chi,
\label{SSA}
\end{equation}
where $\chi$ denotes the common value of the dilaton fluctuation 
either in the string or in the Einstein frame\footnote{This result is  true 
in four space-time dimensions. In higher dimensions the Einstein frame dilaton and its fluctuation are
redefined as $\varphi_{e} = \sqrt{2/(d -1)} \varphi_{s}$ and as $\chi_{e} = \sqrt{2/(d -1)} \chi_{s}$ where $d$ denotes the 
number of spatial dimensions ( $d=3$ in the case discussed here).}; in Eq. (\ref{SSA}) $\delta_{s}$ denotes  the 
scalar fluctuation of the corresponding entry of the metric tensor.
Recalling the conventional decomposition of the scalar fluctuations in the string frame (see e.g. Eq. (\ref{S1})) and analogously in the Einstein frame,  
Eq. (\ref{SSA}) implies the following set of relations between the perturbed entries of the metric in the 
two frames:
\begin{equation}
\phi_{s} = \phi_{e} + \frac{\chi}{2},\qquad \psi_{s} = \psi_{e} - \frac{\chi}{2}, \qquad E_{s} = E_{e}= E, \qquad B_{s} = B_{e} = B,
\label{rel1}
\end{equation}
where $E$ and $B$ denotes the common value of the corresponding fluctuations either in the string or in the Einstein frame.
The gauge-invariant curvature fluctuation in String and Einstein frames are defined as
\begin{equation}
{\cal R}_{e} = - \psi_{e} - \frac{{\cal H}_{e}}{\varphi'}\chi ,\qquad 
{\cal R}_{s} = - \psi_{s} - \frac{{\cal H}_{s}}{\varphi'}\chi.
\label{def1}
\end{equation}
Equation (\ref{def1}) seems to imply that ${\mathcal R}_{e} \neq {\mathcal R}_{s}$: 
on the one hand $\psi_{e} \neq \psi_{s}$ and, on the other hand,  Eq. (\ref{transition}) implies that  ${\mathcal H}_{e} \neq {\mathcal H}_{s}$. However, 
the mismatch between $\psi_{e}$ and $\psi_{s}$ is exactly compensated by 
 $({\mathcal H}_{e} - {\mathcal H}_{s})= - \varphi^{\prime}/2$ so that, eventually, Eqs. (\ref{def1}) and (\ref{rel1}) imply
\begin{equation}
{\mathcal R}_{s} = {\mathcal R}_{e}= {\mathcal R},
\label{def2}
\end{equation}
where ${\mathcal R}$ denotes the common value of the curvature perturbations on comoving 
orthogonal hypersurfaces in the two conformally related frames. 
In the uniform dilaton gauge (i.e. $\chi =0$) we have that ${\mathcal R} = - \psi$ where 
$\psi$ denotes the common value of the longitudinal degree of freedom of the metric since,
in this case, $\psi_{e} = \psi_{s} = \psi$. All in all we can say that the curvature perturbations 
on comoving orthogonal hypersurfaces are both gauge-invariant 
and frame-invariant.
Unlike curvature perturbations, the Bardeen potential are gauge-invariant but not frame-invariant\footnote{In general terms the two Bardeen potentials and the 
dilaton fluctuation are defined, respectively, as $\Phi= \phi + \left[ (B-E^{\prime}) a\right]^{\prime}/a$, $\Psi=\psi -{\mathcal H}(B-E^{\prime})$
and $X= \chi + \varphi^{\prime} (B-E^{\prime})$.} 
\begin{eqnarray}
&&
\Phi_e= \Phi_s - {\chi \over 2} -{\varphi' \over 2}(B-E')
=\Phi_s -\frac{1}{2} X_s,
\nonumber\\
&&
\Psi_e= \Psi_s + {\chi \over 2} +{\varphi' \over 2}(B-E')
=\Psi_s + \frac{1}{2} X_s,
\label{fin}
\end{eqnarray}
while, as anticipated, $X_e= X_s$ and ${\mathcal R}_{e} = {\mathcal R}_{s}$. Equation (\ref{fin}) demonstrate that the Bardeen potential are not frame-invariant 
and are therefore not the best quantities to analyze when discussing the fluctuations of this model. Still from Eq. (\ref{fin}) we might notice that the combination 
$(\Phi_{e} + \Psi_e)= (\Phi_{s} + \Psi_{s})$ is both gauge-invariant and frame-invariant. 
\subsection{Frame transformations for the background}
We shall now apply the transformation (\ref{RR1}) to the evolution equations of the background in different coordinate 
systems. Since the conformal time coordinates are frame-invariant,   to perform swiftly the correct 
transition from string to the Einstein frames, it is appropriate to rewrite the system of Eqs. (\ref{b1})--(\ref{b2}) 
in the conformal time coordinate, i.e. $a_{s}(\tau) d\tau = d t$ where $\tau_{e} = \tau_{s} = \tau$ 
is the common value of the conformal time coordinate either in the string or in the Einstein frame and 
$t$ is the cosmic time coordinate in the string frame. 
After this coordinate change, Eqs. (\ref{b1})--(\ref{b2}) are
\begin{eqnarray}
&& {\overline{\varphi}'}^2  = 3 {\mathcal H}_{s}^2 + V a_{s}^2 ,\qquad {\mathcal H}_{s}' - {\mathcal H}_{s}^2 = {\mathcal H}_{s} \overline{\varphi}',
\label{back1}\\
&& {\overline{\varphi}'}^2  - 2 \overline{\varphi}'' + 2 {\mathcal H}_{s} \overline{\varphi}' +
3 {\mathcal H}_{s}^2 + \frac{\partial V}{\partial \overline{\varphi}} \,a_{s}^2 - V\,a_{s}^2 =0.
\label{back2}
\end{eqnarray}
Equations (\ref{back1}) and (\ref{back2}) can be phrased directly in terms of $\varphi$ by recalling that
$\overline{\varphi}^{\prime} = \varphi^{\prime} - 3 {\mathcal H}_{s}$:
\begin{eqnarray}
&& {\varphi^{\prime}}^2 + 6 {\cal H}_{s}^2 - 6 {\mathcal H}_{s} \varphi^{\prime} = V a_{s}^2 ,\qquad {\mathcal H}_{s}' = {\mathcal H}_{s} \varphi' - 2 {\mathcal H}_{s}^2,
\label{b2in}\\
&& 2 \varphi^{\prime\prime} + 4 {\cal H}_{s} \varphi^{\prime} - 6 {\mathcal H}_{s}' - 6 {\cal H}_{s}^2 + V a_{s}^2- \frac{\partial V}{\partial\overline{\varphi}} a_{s}^2 =0,
\label{b3in}
\end{eqnarray}
where ${\mathcal H}_{s} = a_{s}'/a_{s}$ and the prime denotes derivation with respect to $\tau_{s}=\tau_{e} = \tau$.
The dilaton and the conformal time coordinate do not transform under conformal rescaling so that $\tau$ and $\varphi$ denote the common values of the corresponding variables in both frames.
With this observation by inserting Eq. (\ref{transition}) into  Eqs. (\ref{b2in}) and (\ref{b3in}) the explicit form of Eqs. (\ref{Eone}), (\ref{Etwo}) and (\ref{Ethree}) appearing 
in section \ref{sec2} is readily obtained. 

\subsection{Frame transformation for the fluctuations}

Equations (\ref{II}), (\ref{III}) and (\ref{IV})  account for the evolution of the scalar modes 
in the string frame in the uniform dilaton gauge. One of the virtues 
of this coordinate system is that the corresponding Einstein frame equations  follow from 
Eqs. (\ref{II}), (\ref{III}) and (\ref{IV}) by transforming the background according to Eq. (\ref{transition}).
Therefore, using Eq. (\ref{transition}) into Eqs. (\ref{II}), (\ref{III}) and (\ref{IV}) we obtain without problems the following set 
of equations
\begin{eqnarray}
&& E'' + 2 {\cal H}_{e} E' + \psi - \phi=0,
\label{Ie}\\
&&  \psi' + {\cal H}_{e} \phi=0,
\label{IIe}\\
&& [\psi'' + 2 {\cal H}_{e} \psi'] + {\cal H}_{e} \phi' + ({\cal H}_{e}' + 2 {\cal H}_{e}^2) \phi =0,
\label{IIIe}\\
&& - 6 {\cal H}_{e}\psi' - V a^2_{e} e^{\varphi} \phi + 2 \nabla^2 [\psi + {\cal H}_{e} E'] =0,
\label{IVe}
\end{eqnarray}
where, as already mentioned, $\phi$, $\psi$ and $E$ denote the common values of the metric fluctuations 
either in the string or in the Einstein frame and the conformal time coordinate is unaltered in the transition 
between the two frames.
\end{appendix}


\newpage 

\begin{thebibliography}{999}

\bibitem{WMAP1} D.~N.~Spergel {\it et al.},   Astrophys.\ J.\ Suppl.\  {\bf 148}, 175 (2003); D.~N.~Spergel {\it et al.},
{\em ibid.} \ {\bf 170}, 377 (2007);  L.~Page {\it et al.} {\em ibid.}  {\bf 170}, 335 (2007).

\bibitem{WMAP2} B.~Gold {\it et al.},  Astrophys.\ J.\ Suppl.\ {\bf 192}, 15 (2011); 
D.~Larson,  {\it et al.},  {\em ibid.}  {\bf 192}, 16 (2011); C.~L.~Bennett {\it et al.}, {\em ibid.}\ {\bf 192}, 17 
(2011); G.~Hinshaw {\it et al.},  {\em ibid.} {\bf 208} 19 (2013); C.~L.~Bennett {\it et al.},   {\em ibid.} {\bf 208} 20 (2013); 
 P.~A.~R.~Ade {\it et al.}  [Planck Collaboration],   Astron.\ Astrophys.\  {\bf 571}, A22 (2014); Astron.\ Astrophys.\  {\bf 571}, A16 (2014);
  P.~A.~R.~Ade {\it et al.}  [Planck Collaboration],  arXiv:1502.02114 [astro-ph.CO].

\bibitem{inf1} A.H. Guth, Phys. Rev. D {\bf 23}  347 (1981);
A.D. Linde, Phys. Lett. B {\bf 108}, 389 (1982); A. Albrecht, P.J. Steinhardt, Phys. Rev. Lett. {\bf 48}, 1220 (1982); 
A. Ijjasa, P.  Steinhardt,  A. Loeb, Phys. Lett.  B {\bf 723}, 261 (2013).

\bibitem{inf2} D.~S.~Salopek and J.~M.~Stewart,
  Class.\ Quant.\ Grav.\  {\bf 9}, 1943 (1992); J.~Parry, D.~S.~Salopek and J.~M.~Stewart,  Phys.\ Rev.\ D {\bf 49}, 2872 (1994);
  K. Tomita, Prog. Theor. Phys.  {\bf 67}, 1076 (1982); Phys. Rev. D {\bf 48}, 5634 (1993); N. Deruelle and K. Tomita, Phys. Rev. D {\bf 50}, 7216 (1994);
N. Deruelle and D. Goldwirth, Phys. Rev. D {\bf 51}, 1563 (1995);  M.~Giovannini,
  Phys.\ Lett.\ B {\bf 746}, 159 (2015).

\bibitem{boun1}   M.~Novello and S.~E.~P.~Bergliaffa,
  Phys.\ Rept.\  {\bf 463}, 127 (2008); R.~Brandenberger and P.~Peter,
  arXiv:1603.05834 [hep-th].
 
\bibitem{GI1}  T.~Qiu and Y.~T.~Wang, JHEP {\bf 1504}, 130 (2015); Y.~Wan, T.~Qiu, F.~P.~Huang, Y.~F.~Cai, H.~Li and X.~Zhang,
  JCAP {\bf 1512}, no. 12, 019 (2015); L.~Battarra, M.~Koehn, J.~L.~Lehners and B.~A.~Ovrut, JCAP {\bf 1407}, 007 (2014).

\bibitem{GI2}  V.~A.~Rubakov,  Phys.\ Usp.\  {\bf 57}, 128 (2014); 
  [Usp.\ Fiz.\ Nauk {\bf 184}, no. 2, 137 (2014)]; M.~Libanov, S.~Mironov and V.~Rubakov,  JCAP {\bf 1608}, no. 08, 037 (2016); 
  Y.~Cai, Y.~Wan, H.~G.~Li, T.~Qiu and Y.~S.~Piao,
  arXiv:1610.03400 [gr-qc];  T. Kobayashi, Phys.Rev. D {\bf 94}, 043511 (2016);  Y.~Cai, H.~G.~Li, T.~Qiu and Y.~S.~Piao,
  arXiv:1701.04330 [gr-qc].

\bibitem{nonloc1}   M.~Giovannini, Phys.\ Rev.\ D {\bf 70}, 103509 (2004); Class.\ Quant.\ Grav.\  {\bf 21}, 4209 (2004);
M.~Gasperini, M.~Giovannini and G.~Veneziano, Nucl.\ Phys.\ B {\bf 694}, 206 (2004);
 Phys.\ Lett.\ B {\bf 569}, 113 (2003).

\bibitem{BR} L. H. Ford, Phys. Rev. D {\bf 35}, 2955 (1987); L. Parker, Nature {\bf 261}, 20 (1976); Phys. Rev. {\bf 183}, 1057 (1969); 
L. P. Grishchuk, Ann. (N.Y.) Acad. Sci. {\bf 302}, 439 (1977); 
V. N. Lukash and A. A. Starobinsky, Sov. Phys. JETP {\bf 39}, 742 
 (1974)  [ Zh. Eksp. Teor. Fiz. {\bf 66}, 1515 (1974)].

\bibitem{double1}  C.~Hull and B.~Zwiebach,   JHEP {\bf 0909}, 099 (2009);  JHEP {\bf 0909}, 090 (2009); 
O.~Hohm, C.~Hull and B.~Zwiebach,     JHEP {\bf 1007}, 016 (2010); JHEP {\bf 1008}, 008 (2010)  

\bibitem{double2} 
  W.~Siegel,  Phys.\ Rev.\ D {\bf 47}, 5453 (1993);  Phys.\ Rev.\ D {\bf 48}, 2826 (1993);   A.~A.~Tseytlin,
    Phys.\ Lett.\ B {\bf 242}, 163 (1990);  Nucl.\ Phys.\ B {\bf 350}, 395 (1991);
  M.~J.~Duff,  Nucl.\ Phys.\ B {\bf 335}, 610 (1990).

\bibitem{double3}    H.~Wu and H.~Yang, JCAP {\bf 1407}, 024 (2014); C.~T.~Ma and C.~M.~Shen,
  Fortsch.\ Phys.\  {\bf 62}, 921 (2014); K.~Lee and J.~H.~Park,  Nucl.\ Phys.\ B {\bf 880}, 134 (2014).

 \bibitem{nonloc2} M.~Giovannini,  Class.\ Quant.\ Grav.\  {\bf 22}, 2201 (2005);  Phys.\ Rev.\ D {\bf 55}, 595 (1997);  
 M. Den, Prog. Theor. Phys. {\bf 77}, 653 (1987); K. Tomita and M. Den, Phys. Rev. D {\bf 34}, 3570 (1986);  R. Abbott, 
 B. Bednarz, and S. D. Ellis, Phys. Rev. D {\bf 33}, 2147 (1986);  R. Abbott, S. Barr, and S. Ellis, Phys. Rev. D {\bf 30},720 (1984).

\bibitem{lukash} V.~N.~Lukash,  Sov.\ Phys.\ JETP {\bf 52}, 807 (1980) [Zh. Eksp. Teor. Fiz. {\bf 79}, 1601 (1980)]; 
V.~Strokov,  Astron.\ Rep.\  {\bf 51}, 431-434 (2007); V. N. Lukash and I. D. Novikov, {\it Lectures on the very early universe} in {\it Observational and Physocal Cosmology}, II Canary Islands Winter School of Astrophysics, eds. F. Sanchez, M. Collados and R. Rebolo (Cambridge University Press, Cambridge UK, 1992), p. 3.

\bibitem{hw1} J~-c.~Hwang, Astrophys. J.  {\bf 375}, 443 (1990); Class.\ Quant.\ Grav.\  {\bf 11}, 2305 (1994);
J.~-c.~Hwang and H.~Noh,  Phys.\ Lett.\ B {\bf 495}, 277 (2000).

\bibitem{hw2} J.~-c.~Hwang and H.~Noh, Phys.\ Rev.\ D {\bf 65}, 124010 (2002);  Class.\ Quant.\ Grav.\  {\bf 19}, 527 (2002);
Phys.\ Rev.\ D {\bf 73}, 044021 (2006); H. S. Kim,  and J.~-c.~Hwang  Phys.\ Rev.\ D {\bf 74}, 043501 (2007).

\bibitem{bard1} J. Bardeen, Phys. Rev. {\bf D22}, 1882 (1980); J. Bardeen, P. Steinhardt, and M. Turner, Phys. Rev. {\bf D28}, 679 (1983); 
J.~A.~Frieman and M.~S.~Turner, Phys.\ Rev.\ D {\bf 30}, 265 (1984); R.~H.~Brandenberger, R.~Kahn and W.~H.~Press,
  Phys.\ Rev.\ D {\bf 28}, 1809 (1983);  R.~H.~Brandenberger and R.~Kahn,  Phys.\ Rev.\ D {\bf 29}, 2172 (1984); 

\bibitem{lif} E.~M.~Lifshitz and I.~M.~Khalatnikov,
  Adv.\ Phys.\  {\bf 12}, 185 (1963); E.~M.~Lifshitz Zh. Eksp. Teor. Fiz. {\bf 16}, 587 (1946).

\bibitem{bard2}  H.~Kodama, M.~Sasaki,  Prog.\ Theor.\ Phys.\ Suppl.\  {\bf 78}, 1-166 (1984); 
M. Sasaki, Prog. Teor. Phys. {\bf 76}, 1036 (1986); G.~V.~Chibisov, V.~F.~Mukhanov,  Mon.\ Not.\ Roy.\ Astron.\ Soc.\  {\bf 200}, 535 (1982); 
V.~F.~Mukhanov,  Sov.\ Phys.\ JETP {\bf 67}, 1297 (1988)  [Zh. Eksp. Teor. Fiz. {\bf 94}, 1 (1988)].

\bibitem{wein} S.~Weinberg, Phys.\ Rev.\ D {\bf 77}, 123541 (2008);  E. Elizalde, A. Jacksenaev, S. D. Odintsov, and I. L. Shapiro, Phys. Lett. B {\bf 328}, 297 (1994); 
Class. Quant. Grav. {\bf 12}, 1385 (1995).
  
\end{thebibliography}
\end{document}